\documentclass[12pt]{article}
\usepackage{caption}
\usepackage{subcaption}
\usepackage{booktabs}
\usepackage{graphicx}
\usepackage{tikz}
\usepackage{algorithm, algpseudocodex}
\usepackage{multirow, multicol}
\RequirePackage[authoryear]{natbib}
\usepackage{url} 
\usepackage{hyperref}
\hypersetup{
    hyperindex=true,
    linktocpage = true,
    colorlinks=true,
    linkcolor=blue,
    filecolor=magenta,      
    urlcolor=cyan,
    citecolor=blue,
    pdftitle={Overleaf Example},
    pdfpagemode=FullScreen,
}

\usepackage{amsmath, amsfonts, amssymb, amsthm}

\newcommand{\bi}{\begin{itemize}}
\newcommand{\ei}{\end{itemize}}

\newcommand{\be}{\begin{enumerate}}
\newcommand{\ee}{\end{enumerate}}

\newcommand{\EE}{\mathrm{E}}
\newcommand{\var}{\mathrm{var}}
\newcommand{\cov}{\mathrm{cov}}

\newcommand{\Fcal}{\mathcal{F}}

\newcommand{\Ycal}{\mathcal{Y}}

\newcommand{\0}{\mathbf{0}}

\newcommand{\E}{\mathbf{E}}

\newcommand{\N}{\mathbf{N}}
\newcommand{\T}{\mathbf{T}}

\newcommand{\V}{\mathbf{V}}

\newcommand{\X}{\mathbf{X}}

\newcommand{\w}{\mathbf{w}}
\newcommand{\x}{\mathbf{x}}
\newcommand{\y}{\mathbf{y}}
\newcommand{\z}{\mathbf{z}}

\newcommand{\alphabf}{\boldsymbol{\alpha}}
\newcommand{\betabf}{\boldsymbol{\beta}}

\newcommand{\psibf}{\boldsymbol{\psi}}

\newcommand{\mubf}{\boldsymbol{\mu}}

\newcommand{\xibf}{\boldsymbol{\xi}}

\newcommand{\thetabf}{\boldsymbol{\theta}}

\newcommand{\etabf}{\boldsymbol{\eta}}

\newcommand{\sigmabf}{\boldsymbol{\sigma}}
\newtheorem{definition}{Definition}[section]

\newcommand{\blind}{0}

\begin{document}

\def\spacingset#1{\renewcommand{\baselinestretch}%
{#1}\small\normalsize} \spacingset{1}


\if0\blind
{
  \title{\bf Markov Random Fields: Structural Properties, Phase Transition, and Response Function Analysis}
  \author{J. Brandon Carter\thanks{
    The authors gratefully acknowledge support from the Eunice Kennedy Shriver National Institute on Child Health and Human Development (R01HD088545, R01HD113259, and P2CHD042849) and the National Science Foundation (2502935)}\hspace{1cm}\\[.1cm]
    Department of Statiscal Science \\ Duke University \\[.2cm]
    and \\[.2cm]
    Catherine A. Calder\footnote{calder@austin.utexas.edu} \\[.1cm]
    Department of Statistics and Data Sciences \\ University of Texas at Austin}
  \maketitle
} \fi

\if1\blind
{
  \bigskip
  \bigskip
  \bigskip
  \begin{center}
    {\LARGE\bf Markov Random Fields: Structural Properties, Phase Transition, and Response Function Analysis}
\end{center}
  \medskip
} \fi

\bigskip
\begin{abstract}
This paper presents a focused review of Markov random fields (MRFs)--commonly used probabilistic representations of spatial dependence in discrete spatial domains--for categorical data, with an emphasis on models for binary-valued observations or latent variables. We examine core structural properties of these models, including clique factorization, conditional independence, and the role of neighborhood structures. We also discuss the phenomenon of phase transition and its implications for statistical model specification and inference.
A central contribution of this review is the use of \textit{response functions}, a unifying tool we introduce for prior analysis that provides insight into how different formulations of MRFs influence implied marginal and joint distributions.
We illustrate these concepts through a case study of direct-data MRF models with covariates, highlighting how different formulations encode dependence.
While our focus is on binary fields, the principles outlined here extend naturally to more complex categorical MRFs and we draw connections to these higher-dimensional modeling scenarios.
This review provides both theoretical grounding and practical tools for interpreting and extending MRF-based models.
\end{abstract}

\noindent%
\vfill

\newpage
\spacingset{1.75} 
\section{Introduction}

Markov random fields (MRFs) are a popular class of probability distributions used to model dependencies between discrete-valued variables associated with a finite number of units of analysis. 
As MRFs are commonly applied in an explicitly spatial setting, without loss of generality, we refer to the unit of analysis as an areal unit.
The areal units could be administrative districts or a regular gridded partition of the study domain (e.g., a lattice).
In this paper, we focus our attention primarily on binary MRFs and draw connections to MRFs for categorical discrete observations which arise as natural extensions of the models for binary data.
The application of MRFs for discrete-valued, dependent variables have been used to solve a wide range of inferential and predictive tasks including image segmentation \citep{zhang_etal2001}, tissue classification \citep{feng_etal2012}, remote sensing \citep{moores_etal2020}, species distribution mapping \citep{hogmander_moller1995}, and disease mapping \citep{green_richardson2002}.
In all of these applications, the unifying supposition of the statistical analysis is that areal units which are closer together in space will have similar values; that is, we expect some degree of spatial clustering in the discrete-valued variables of an analysis.
As the set areal units form a partition of the spatial domain -- a discrete set of locations -- the ``closeness'' of the areal units is not defined by a traditional distance metric, but rather the dependence between areal units is represented through an \textit{natural undirected graph} (NUG) comprised of vertices, corresponding to the areal units, and undirected edges between pairs of vertices, which indicate mutual dependence between the vertices. 
We follow the terminology of \citet{carter2025mixture} and call undirected graph ``natural'' because it is not an object of inference and is defined \textit{a priori} following one of two common definitions of dependence between areal units (see Section~\ref{sec:discrete_mrfs}).
Mutually dependent vertices and their corresponding areal units are called \textit{neighbors} and the undirected graph is sometimes referred to as a \textit{neighborhood} or neighborhood structure, historical terminology, to which the ``N'' in the acronym NUG signals.
The pairwise dependence relationships, defined \textit{a priori}, are most commonly determined by the spatial contiguity of the areal units (e.g., areal units in a 2-dimensional setting that share a border are designated as neighbors).
The appeal of an MRF is that the dependencies represented in the NUG can be perfectly represented by the MRF probability distribution.
In fact, an MRF can be defined as a probability distribution whose conditional independence relationships define a NUG \citep{cressie1993}. 

Due to the historical development of MRF models in statistical mechanics and their later adoption as a modeling tool in statistics, there is an inconsistent application of terminology to define different MRF \textit{formulations}.
In this paper, we use formulation to mean the mathematical specification of the MRF probability distribution itself and distinguish modeling categories or scenarios by the overall structure of the probability model, of which the MRF may be a key or the sole component.  
We aim to provide a comprehensive comparison of the distinct MRF formulations commonly used to specify models for spatially-dependent, binary data, and draw connections to models for categorical data.
Additionally, we address a lack of clarity about the phenomenon known as \textit{phase transition}, exhibited in some MRF formulations, and consequent modeling choices due to phase transition. 
In order to provide better understanding of common formulations of MRF-based statistical models that are used in the variety of analysis tasks highlighted above, we introduce a new prior analysis tool, \textit{response functions}, which we use to highlight different properties of the distinct MRF formulations, including that of phase transition. 
As discussed in detail in Section~\ref{sec:response_functions}, the concept of response functions is derived from the statistical mechanics approach to understand the real phenomena of phase transition in physical models; we expand on this approach by formalizing response functions as a prior-analysis method which can help characterize complex probability distributions.
Using response functions, we demonstrate properties, both desirable and undesirable, of the MRF models. 

To begin our classification of different MRF formulations and their application to different modeling scenarios, we introduce some notation.
First, let $q$ denote an MRF probability distribution for binary or categorical variables and $p$ be any other probability distribution.
We let $\y$ denote a vector of observed variables, $\z$ a vector latent variables,  and $\X$ a matrix of covariates.
Broadly, we can classify MRF-based statistical models into three types.
First, we have direct data models, $q(\y)$, where our observed variables are discrete and spatial dependence is modeled directly on the observed variables.
These direct data models, such as the autologistic model of \citet{besag1974}, are common in species distribution modeling \citep{heikkinen_hogmander1994, hogmander_moller1995, hughes_etal2011} and often incorporate dependence on observed covariates.
Hidden Markov random fields (HMRFs) models are a second type.  HMRFs can be viewed as a type of generative model -- models that describe a data-generating process. In an HMRF we model spatial dependence on a latent, discrete-valued variable, $q(\z)$, and then model the observed variables as conditionally independent given the latent variable, $p(\y|\z)=\prod_{i=1}^np(y_i|z_i)$, leading to a joint model $p(\y,\z)=p(\y|\z)q(\z)$. 
These models are commonly used in the image segmentation setting  \citep{li2009markov, winkler2006image}
Lastly, we consider conditional random fields (CRFs) -- a type of discriminative model -- which model the conditional distribution of latent states $\z$ as an MRF, $q(\z|\y)$, given observed variables $\y$, the distribution of which need not be known \citep{lafferty2001conditional}.
Here, our characterization of $\z$ as latent variables is a bit loose as $\z$ will be observed during model training and unobserved for prediction tasks.
CRFs have been applied to a wide range of machine learning tasks from natural language processing to image segmentation to recognition and labeling \citep{yu2020comprehensive}.
The explicit modeling of the conditional distribution is useful to distinguish CRFs from the HMRF setting, but as we will see, the CRF formulaically is similar to a direct-data model with covariates.


While all three of these categories of models that use an MRF for spatial dependence appear to be quite different, the formulations and mathematical properties of each category fit coherently within the general MRF framework.
In Section~\ref{sec:discrete_mrfs}, we highlight the most general formulation which encompasses all model specifications used in the three  categories of MRF models listed above.
From this general formulation of an MRF, we highlight the three most prevalent formulations for binary data, which are unevenly favored for specification of an MRF in the above model categories.
We also introduce a common formulation for categorical data, which arise as the natural extension from the binary formulations. 
In order to better understand the properties of these different MRF formulations, the key distinguishing feature of MRF behavior,
we introduce our new prior analysis tool, response functions, in Section~\ref{sec:response_functions}.
We provide a clear definition of phase transition in Section~\ref{sec:phase_transition} and highlight key statistical properties of phase transition that can be characterized by different response functions. 
With the property of phase transition understood, we explain modeling implications and provide recommendations.  
Across Sections~\ref{sec:external_field} and \ref{sec:pairwise_clique_potential}, we show how the three categories of models with an MRF component are all related within the general MRF formulation, provide extensions to the formulations for categorical data and
as a special case study, we demonstrate an undesirable property of the centered-autologistic model in Section~\ref{sec:covariate_external_field}.
We provide a cursory review model fitting methods in Section~\ref{sec:methods_fitting} and offer concluding remarks in Section~\ref{sec:discussion}.

\section{Discrete Markov Random Fields}
\label{sec:discrete_mrfs}
In this section, we specify the most general form of an MRF probability distribution, narrow this specification to a general exponential family parameterization, and review the key statistical properties of exponential families.
These properties of exponential families will be key to understanding how spatial dependence is encoded in an MRF. 
From the general exponential family specification, we then enumerate the most commonly used formulations of the MRF for binary and categorical data.
%

We proceed by defining a MRF, $q(\y|\xibf)$, 
where $\xibf$ is the set of parameters which govern the MRF.
Whether the areal units are geographic regions or represent a different spatial system, a defining feature of an MRF is that the spatial dependence relationships can also be represented by an NUG. 
For $n$ areal units, let $\N=\{\V,\E\}$ be a NUG which defines the dependence relationships between areal units.
Define $\V=\{1,\dots,n\}$ to be the set of vertices corresponding to the areal units and $\E$ to be the set of undirected edges $\{i,j\}$ which define the mutual dependence between areal units.
Vertices $i$ and $j$ are called neighbors if and only if there is an undirected edge between them.
We also define 
$$\partial(i)=\{j:\{i,j\}\in\E\}$$
to be the set of all neighbors of $i$.
The particular assignment of discrete values, $\y$, associated with the vertices of the graph is called a \textit{configuration} of the graph.
Thus, a realization from an MRF probability distribution provides a configuration of the graph. 
Let $\Ycal$ be the sample space of $\y$, that is the set of all possible configurations.

\begin{figure}[ht]
\centering
\begin{subfigure}[b]{0.3\textwidth}
\centering
\begin{tikzpicture}
        \def\nx{3} \def\ny{3}
        \draw [black] (0,0) grid (\nx,\ny);
    \end{tikzpicture}
\caption{}
\label{fig:lattice}
\end{subfigure}
\begin{subfigure}[b]{0.3\textwidth}
\centering
\begin{tikzpicture}[>=stealth, thick, node distance=12mm, main/.style = {draw, circle}] 
\node[main] (1) {}; 
\node[main] (2) [below of=1] {}; 
\node[main] (3) [below of=2] {};
\node[main] (4) [right of=1] {}; 
\node[main] (5) [right of=2] {}; 
\node[main] (6) [right of=3] {};
\node[main] (7) [right of=4] {}; 
\node[main] (8) [right of=5] {}; 
\node[main] (9) [right of=6] {};
\draw (1) -- (2);
\draw (1) -- (4);
\draw (2) -- (3);
\draw (2) -- (5);
\draw (3) -- (6);
\draw (4) -- (5);
\draw (4) -- (7);
\draw (5) -- (6);
\draw (5) -- (8);
\draw (6) -- (9);
\draw (7) -- (8);
\draw (8) -- (9);
\end{tikzpicture} 
\caption{}
\label{fig:rook}
\end{subfigure}
\begin{subfigure}[b]{0.3\textwidth}
\centering
\begin{tikzpicture}[>=stealth, thick, node distance=12mm, main/.style = {draw, circle}] 
\node[main] (1) {}; 
\node[main] (2) [below of=1] {}; 
\node[main] (3) [below of=2] {};
\node[main] (4) [right of=1] {}; 
\node[main] (5) [right of=2] {}; 
\node[main] (6) [right of=3] {};
\node[main] (7) [right of=4] {}; 
\node[main] (8) [right of=5] {}; 
\node[main] (9) [right of=6] {};
\draw (1) -- (2);
\draw (1) -- (4);
\draw (1) -- (5);
\draw (2) -- (3);
\draw (2) -- (5);
\draw (2) -- (4);
\draw (2) -- (6);
\draw (3) -- (6);
\draw (3) -- (5);
\draw (4) -- (5);
\draw (4) -- (7);
\draw (4) -- (8);
\draw (5) -- (6);
\draw (5) -- (8);
\draw (5) -- (7);
\draw (5) -- (9);
\draw (6) -- (9);
\draw (6) -- (8);
\draw (7) -- (8);
\draw (8) -- (9);
\end{tikzpicture} 
\caption{}
\label{fig:queen}
\end{subfigure}
\caption{Example of a regular lattice and the NUGs  that can be used to represent dependence between the areal units of the lattice.}
\label{fig:ex_lattice}
\end{figure}
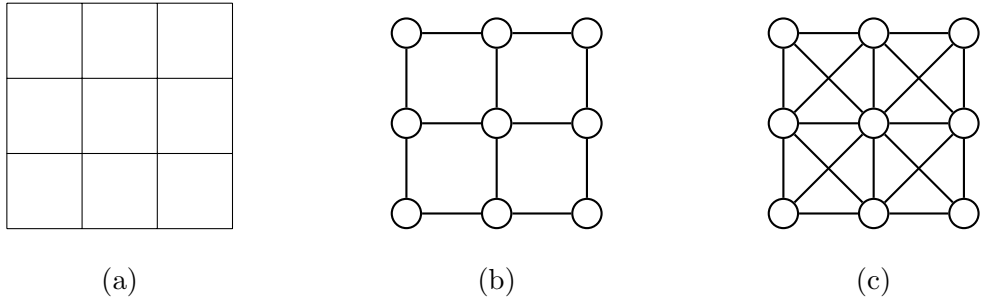

The NUG in the discrete spatial setting generally takes one of two forms, corresponding to either a \textit{first} or \textit{second-order} dependence structure, where neighbor relationships are defined by the contiguity of areal units. 
We reemphasize that this is the meaning implied by \textit{natural} in ``natural undirected graph,'' -- the default specification of the NUG as one of these two structures characterizing the spatial dependence between areal units. 
A NUG based on \textit{first-order} dependence is defined such that the set of neighbors of areal unit $i$ are the units that have a border touching at more than just a point.
In a regular lattice, this corresponds to the areal units directly above, below, left, and right of $i$.
Figure~\ref{fig:lattice} depicts a three-by-three regular lattice with the corresponding NUG defined by a first-order dependence in Figure~\ref{fig:rook}.
A NUG based on \textit{second-order} dependence defines neighbors to be the areal units that have common border or have corners that touch, giving a set of eight neighbors for any areal unit in a regular lattice that is not on the border.
Figure~\ref{fig:queen} shows the implied NUG of a second-order dependence for the three-by-three grid. 

We now formalize the definition of an MRF in relation to the NUG, $\N$.
For sets of variables, $\y_A$, $\y_B$, and $\y_C$, indexed by the sets of vertices $A$, $B$, and $C$, let $I(\N)$ be the set of conditional dependence relationships of the form $A\perp B|C$ asserted by $\N$ and $I(q)$, be the set of conditional independence relationships asserted by $q$.
That is, if we can factorize $q$ as $q(\y_A,\y_B|\y_C)=q(\y_A|\y_C)q(\y_B|\y_C)$, then $A\perp B|C\in I(q)$.
A probability distribution $q(\y|\xibf)$ with respect to $\N$ is called an MRF if $I(q)=I(\N)$ for all $\xibf$.
In other words, the NUG for which $q$ is defined perfectly represents the conditional independence relationships between the variables in the MRF.\footnote{Some texts provide a weaker definition, such that $q(\y|\xibf)$ is called an MRF if the full-conditional distributions of $q(\y|\xibf)$ imply the same conditional independence relationships as the NUG \citep[pg. 194]{rue_held2010}.}

NUGs, and consequently MRFs, satisfy three Markov properties: the global, pairwise, and local. 
The global property relies on the notion of a path; a path is any sequence of connected edges in which the sequence of connected vertices are unique.
The global Markov property states that any two sets of vertices, $A$ and $B$, are independent given $C$ if every path from a vertex in $A$ to a vertex in $B$ goes through $C$. 
By the pairwise Markov property, any two non adjacent (non-neighboring) vertices are independent given all other vertices.  
Of particular interest is the local Markov property, which states that vertex $i$ is independent of all other vertices given its neighbors, $\partial(i)$. 
All of the three properties can be used as a starting point to prove the other two \citep[pg 119]{koller_friedman2009}.
From the local Markov property, the full-conditional distributions of an MRF reduce to the conditional distribution of $y_i$ given the neighbors of $y_i$, namely $\y_{\partial(i)}$.

While the above definition provides an intuitive characterization of spatial dependence, the Hammersley-Clifford theorem (unpublished, proof given by \citet{besag1974}) provides what form the distribution $q(\y|\xibf)$ with respect to $\N$ must take to be faithful to the graph.
The theorem depends on the notion of a \textit{clique}. 
For an NUG $\N$, a clique, denoted by $c$, is defined to be any single vertex or set of vertices which are all mutually neighbors.
Let $C=\{c_1,\dots,c_m\}$ denote the set of all cliques in the graph $\N$.
In a first-order NUG of a regular lattice, $C$ contains the cliques comprised of singleton vertices and neighboring pairs of vertices, whereas higher-order cliques are admitted for second-order and other NUGs. 

The Hammersley-Clifford theorem states that an MRF, $q(\y|\xibf)$, such that $q(\y|\xibf)>0$ for $\y\in\Ycal$, with respect to an NUG, $\N$, and corresponding cliques, $C$, can be factorized as
\begin{equation}
\label{eq:hc_thm}
q(\y|\xibf) = \frac{1}{Z(\xibf)} \prod_{c\in C} \phi_c(\y_c|\xibf_c),
\end{equation}
where $\phi_c$ are arbitrarily chosen strictly positive functions for each clique $c\in C$ and $\xibf_c$ are the parameters associated with the $c$th clique function. 
The \textit{partition function}, as called in statistical mechanics,
$$
Z(\xibf) = \sum_{\y\in\Ycal}\prod_{c\in C} \phi_c(\y_c|\xibf_c),
$$
is the normalizing constant (i.e., the sum over all possible configurations) and $\xibf$ is the set of all parameters in the clique functions.\footnote{The traditional $Z$ notation for the partition function comes from the German word \textit{Zustandssumme}, meaning sum over states.}
The power of this theorem
comes from the ability to specify an MRF either through the clique functions directly or through full-conditional distributions implied by the graph. 
Equation~\ref{eq:hc_thm} provides the most general form an MRF can take through arbitrary specification of the clique functions, $\phi_c(\y_c|\xibf_c)$.
Through historical precedent and ease of interpretation, an MRF is more commonly specified through $f_c=\log \phi_c$, where the $f_c$'s are called the clique potentials.
There is some discrepancy in the literature with regards to terminology, as ``clique potential'' has been used for both $f_c$ and $\phi_c$. \citet{cressie1993} is perhaps the most precise and prefers the term negpotential, as $-\sum_{c\in C}f_c(\y_c|\xibf)$ has a direct interpretation of the potential energy function (i.e., the Hamiltonian) in statistical mechanics. Following the (majority) precedent in the statistics/machine learning literature, we refer to $f_c$ as clique potential and $\phi_c$ as the clique functions.
In the following, we show how a particular form of the clique potentials results in an exponential family form of the MRF.
In contrast, in Section~\ref{sec:autologistic}, reviews the autologistic formulation of the MRF, a special case of auto models which assume an exponential family parameterization for the full-conditional distributions of an MRF \citep{besag1974}.

\subsection{Exponential Family Form}
\label{sec:exp_fam}
We present what form the clique potentials must take for the MRF to be an exponential family with natural parameters $\xibf$. 
When the clique potentials, $f_c(\y_c|\xi_c)$, can be factorized as $\xi_c g_c(\y_c)$, a function of the data (i.e., the observed configuration) times the parameter associated with the clique, then the MRF follows an exponential family where
\begin{equation*}
\begin{aligned}
    q(\y|\xibf) &= \frac{1}{Z(\xibf)} \exp\left(\sum_{c\in C}\xi_c g_c(\y_c) \right) \\
    &= \frac{1}{Z(\xibf)} \exp\left(\sum_{i=1}^n \xi_i g_i(y_i) + \sum_{\{i,j\}\in C}\xi_{ij}g_{ij}(y_i,y_j) + \sum_{\{i,j,k\}\in C}\xi_{ijk}g_{ijk}(y_i,y_j,y_k) + \dots \right).
\end{aligned}
\end{equation*}
While a more general exponential family form is possible by permitting the natural parameters to be functions of another set of parameters, this particular factorization of the clique potentials encompasses the commonly used cases which we will highlight later on in this section. 

The above equation specifies a unique distribution; however, it is clear that each individual clique potential is a sufficient statistic and in the absence of data replication, as is often the case in spatial applications, the individual $\xi_c$ are not identifiable. 
One reasonable solution is to assume a common parameter, $\xi_c$, and $g_c$ function for each clique size $c=1,2,\dots$ admitted by the graph (e.g., singleton, pairwise, etc.), or possibly a common parameter and $g$ function for each clique orientation (e.g., in a regular lattice with first-order dependence, assume $\xi_{ij}=\psi_1$ for vertically adjacent areal units, $\xi_{ij}=\psi_2$ for horizontally adjacent areal units and $g_{ij}(y_i,y_j)=g(y_i,y_j)$.)
Without loss of generality, let $\xi_i = \xi_1$, $\xi_{ij}=\xi_2$ and so forth for higher-order cliques that are admitted by the graph and assume common $g$ functions for each clique size (e.g., $g_i(y_i)=g(y_i)$ for all $i\in \V$ and $g_{ij}(y_i,y_j)=g(y_i,y_j)$ for all $\{i,j\}\in \E$).
We adopt these assumptions for simplicity of presentation, but, as noted above, the following results can easily be extended to greater granularity in the clique potentials \citep[for example,][]{arnesen_tjelmeland2015}.
We now provide a definition of a simplified exponential family MRF. 
\begin{definition} 
\label{def:sefmrf}
Assume common parameters and $g$ functions for each clique size for an MRF with respect to $\N$. Then we say a probability distribution is a simplified exponential family (SEF) MRF with the form
\begin{equation}
\label{eq:ex_fam_mrf}
q(\y|\xibf) = \frac{1}{Z(\xibf)} \exp\left(\xibf'\T(\y)\right).
\end{equation}
In Equation~\ref{eq:ex_fam_mrf}, $\xibf=[\xi_1,\xi_2,\dots,\xi_K]'$ is the vector of natural parameters,
$$\T(\y) = [T_1(\y), T_2(\y),\dots, T_K(\y)]'$$ is the vector of sufficient statistics for each clique size admitted by the graph, and $K$ is the largest clique size in the graph.
We define
$$
T_k(\y)=\sum_{c\in C,|c|=k}g(\y_c),
$$ 
with $|c|$ denoting the cardinality of the set $c$. 
\end{definition}
From Definition~\ref{def:sefmrf}, $T_k(\y)$ is the sum of the clique potentials for cliques of size/order $k$ and only exists if cliques of order $k$ are admitted by the graph. 
We refer to all a parameters associated with cliques of order $k>1$ as spatial dependence parameters as they affect the mutual dependence between variables.
Notably, for a regular lattice with first-order dependence, the highest order clique admitted by the graph is pairwise. 
Even when higher order cliques are admitted by the graph, it is quite common to specify a model with only pairwise dependence, as do all the common forms we present below.
%


For exponential families, the log partition function has a special relationship to the cumulants of the sufficient statistics.
Letting $A(\xibf)=\log Z(\xibf)$ be the log-partition function, we have 
$$
A(\xibf) = \log\left(\sum_{y\in\Ycal}\exp(\xibf'\T(\y))\right) .
$$
By first taking the derivative of $A(\xibf)$ with respect to $\xibf$ (that is, $\nabla A(\xibf) = [\frac{\partial r(\xibf)}{\partial \xi_1}, \frac{\partial r(\xibf)}{\partial \xi_2}, \dots]'$), we obtain the result that $\nabla A(\xibf) = \EE(\T(\y))$.
Thus, the first derivative of the log partition function is the expected value (i.e., first cumulant) of the vector of sufficient statistics.
Taking the second derivative we obtain the result that $\nabla^2A(\xibf)=\cov(\T(\y))$, where $\nabla^2A(\xibf)$ is the matrix of partial second derivatives. Specifically, we have 
$$
\nabla^2_{ij}A(\xibf) = \cov(T_i(\y),T_j(\y)) = \frac{\partial^2A(\xibf)}{\partial\xi_i\partial\xi_j}.
$$
These mathematical results are particularly useful in characterizing the behavior of MRFs across different values of $\xibf$.
Drawing connections between the statistical mechanics literature and these results, we provide a new prior analysis tool in Section~\ref{sec:response_functions}, \textit{response functions}, which are well suited to profile how different MRF formulations encode spatial dependence.
For example, in Section~\ref{sec:phase_transition}, we show how the well-studied phenomena of phase transition in the Ising model is evidenced in the behavior of $\var(T_2(\y))$ as $\xi_2$ varies \citep{stoehr2017}.



\subsection{Formulations of the Simplified Exponential Family MRF}
\label{sec:special_cases_sefmrf}
In the following, we present the most common formulations of the SEF-MRFs that appear in the literature.
For binary data these are the \textit{physics-Ising} formulation, the \textit{Ising} formulation, and the \textit{autologistic} formulation.
For categorical data, the \textit{Potts} formulation is routinely used; here, we present the standard Potts formulation and note extensions in Section~\ref{sec:pairwise_clique_potential}. 
In our terminology, consistent with how the ``Ising model'' is used in the statistics literature, the Ising formulation is identical to the 2-state Potts formulation, distinct from the physics-Ising model.
The Ising and physics-Ising formulations are, in fact, related via a variable transformation. 
While each of these formulations have precedents in specific modeling settings (e.g., the autologistic formulation has primarily been used as a direct data model), we emphasize that each formulation can be used as a direct data model, a component in a hierarchical framework (e.g., as a prior for a latent discrete variable) or in a more flexible modeling setting (e.g., conditional random fields). 
We continue our presentation of the formulations first as a direct data model for $\y$ and will refer to $q(\y|\xibf)$ as the model for our data.

For all of the formulations, we assume only pairwise dependence, we set $\xi_k=0$ for $k>2$, and let $\xi_2=\psi$.
For the binary formulations, let $\xi_1=\alpha$ and $g(y_i)=y_i$, with a slight modification necessary for the general Potts model as explained below. 
Assuming only pairwise dependence, the SEF-MRF as a direct data model takes the general form
$$
q(\y|\xibf) = \frac{1}{Z(\xibf)} \exp\left(\alpha\sum_i g(y_i) + \psi\sum_{i\sim j} g(y_i,y_j) \right),
$$
where $\xibf=[\alpha, \psi]'$, $\sum_i\equiv\sum_{i=1}^n$, and $i\sim j$ denotes the set of all neighbors in the graph (i.e., the set $\E$). 
Each model is primarily distinguished by the function $g(y_i,y_j)$, and the physics-Ising model is distinguished by the values which $y_i$ takes on. 
Let $I(\cdot)$ be the indicator function that evaluates to one if the statement inside the parentheses is true and zero otherwise.
Table~\ref{tab:mrf_formulations} gives a summary of all the models in terms of $g$ functions and values of $y_i$.

\begin{table}[ht!]
    \centering
    \begin{tabular}{lll}
        Model & $y_i$ & $g(y_i,y_j)$\\
         \hline
        Physics-Ising & $\{-1,1\}$ & $y_iy_j$ \\
        Autologistic & $\{0,1\}$ & $y_iy_j$\\
        Ising & $\{0,1\}$ & $I(y_i=y_j)$ \\
        Potts & $\{0,1,\dots,k-1\}$ & $I(y_i=y_j)$\\
    \end{tabular}
    \caption{Summary table of the common formulations of a MRF for binary and categorical data.}
    \label{tab:mrf_formulations}
\end{table}

The parameters $\alpha$ and $\psi$ can respectively be interpreted through the sufficient statistics $T_1(\y)=\sum_{i} g(y_i)$ and $T_2(\y)=\sum_{i\sim j} g(y_i,y_j)$:  
Let $\y^*$ be an alternate configuration, then we can write the log-odds of configuration $\y^*$ over configuration $\y$ as
\begin{equation}
\label{eq:logodds}
\log\left(\frac{q(\y^*|\xibf)}{q(\y|\xibf)}\right) = \alpha(T_1(\y^*) - T_1(\y)) + \psi(T_2(\y^*) - T_2(\y)),
\end{equation}
which gives a useful pattern to interpret the effect of $\alpha$ and $\psi$ on the configurations of the NUG.
For example, we can compare two different configurations such that $T_1(\y)=T_1(\y^*)$, but differ in $T_2$ to interpret the marginal effect of $\psi$ and vice versa to interpret the marginal effect of $\alpha$. 
Concrete examples are given below as we elaborate on the different MRF formulations and give interpretations of the parameters for each.

\subsubsection{Physics-Ising Formulation}
\label{sec:physics-ising}
We begin with the physics-Ising model \citep{ising1925} due to its pioneering place in history and consequential impact on the fields of physics and statistics \citep{duminil-copin2022}. 
Much of the terminology for MRFs, including that of phase transition, derives from models used in statistical mechanics, of which the physics-Ising model is the most famous. 
The physics-Ising model was first proposed as a model for the behavior of ferromagnetic metals whose atoms are arranged in a lattice.
Accordingly, $y_i$ takes on the values $-1$ and $1$ to represent the down spin or up spin of each atom.
The physics-Ising model is defined as follows.

\begin{definition}
\label{def:physics-ising}
When $g(y_i,y_j)=y_iy_j$ and $g(y_i)=y_i$, the MRF has an physics-Ising formulation given by  
\begin{equation*}
\label{eq:physics-ising}
   q(\y|\xibf) = \frac{1}{Z(\xibf)} \exp\left(\alpha\sum_i y_i + \psi\sum_{i\sim j}y_i y_j \right),     
\end{equation*}
with $\y\in\{-1,1\}^n$.
\end{definition}

When an MRF has the physics-Ising formulation, the full-conditional distributions are 
\begin{equation}
\label{eq:physics-ising_full_cond}
q(y_i|\y_{\partial(i)}) = \frac{\exp\left(y_i\left(\alpha + \psi\sum_{j\in\partial(i)}y_j\right)\right)}{\exp\left(-\alpha - \psi\sum_{j\in\partial(i)}y_j\right) +\exp\left(\alpha + \psi\sum_{j\in\partial(i)}y_j\right)},
\end{equation}
for $y_i\in\{-1,1\}$.
While statistical mechanics favors a slightly different parametrization than the one above, the concepts from statistical mechanics are analogous as the sufficient statistics are the same.
The pairwise sufficient statistic, $T_2(\y)$, is the number of matches, $\sum_{i\sim j}I(y_i = y_j)$, minus the number mismatches, $\sum_{i\sim j}I(y_i \ne y_j)$, for the configuration $\y$.
It follows that $T_2(\y)\in\{t:t=-|\E|+2u,u\in\mathbb{N},u\le|\E|\}$, where $\mathbb{N}$ is the set of natural numbers and $|\E|$ is the number of edges in the NUG. 
In other words, for every unit increase in the number of matches of a configuration, the pairwise sufficient statistic increases by two.
Note that for different graph structures, the sample space of the pairwise sufficient statistic may only be a subset of the above set for $T_2(\y)$.  Similarly, $T_1(\y)\in\{t:t=-|\V|+2u,u\in\mathbb{N},u\le|\V|\}$, which set also describes the sample space of $T_1$.

The spatial dependence parameter $\psi$ (directly related to the inverse temperature in statistical mechanics) governs the preference of the system for neighboring atoms to be matching in their spins (i.e., both 1s or both -1s).
When $\psi>0$, the system is said to be \textit{cooperative}, \textit{attractive}, or \textit{ferromagnetic} and variables at neighboring areal units are more likely to match;  for $\psi=0$, the spins are independent.
When $\psi<0$, the system is said to be \textit{competitive}, \textit{repulsive}, or \textit{antiferromagnetic} such that mismatches between neighboring variables are preferred, a pattern that is uncommon in most applications of spatial statistics, but one that is not inconceivable.  
Consider two configurations $\y^*$ and $\y$, where $\y^*$ has one more pair of matching areal units than $\y$, but that $T_1(\y^*)-T_1(\y)=0$.
The difference, $T_2(\y^*)-T_2(\y)=2$, so from Equation~\ref{eq:logodds} we have that the log-odds of configuration $\y^*$ over $\y$ is equal to $2\psi$.
The same logic applies to interpreting the effect of $\psi$ in the full-conditional distributions when changing a single neighboring areal unit from negative one to one. 

The parameter $\alpha$ is referred to as the \textit{external field}: For positive values of $\alpha$, the system assigns higher probability to configurations with a net up spin (i.e., more 1s than -1s), thus $E(T_1(\y))>0$ for $\alpha>0$, and negative values favor configurations with a net down spin. 
Following a similar procedure as before, assume that the number of matches for two configurations, $\y^*$ and $\y$, are the same ($T_2(\y^*)=T_2(\y)$), but that $\y^*$ has one more areal unit with the value one. 
Then, the log-odds of configuration $\y^*$ over $\y$ is $2\alpha$.

It is important to keep in mind that the space of configurations, $\Ycal$, is not concisely explored by the two scenarios described above, namely, holding $T_1$ constant and incrementing $T_2$ or holding $T_2$ constant and incrementing $T_1$.
Consider a configuration of all negative ones on a regular first-order lattice, changing a single internal areal unit to positive one will increase $T_1$ by two, but decrease $T_2$ by four.

\subsubsection{Autologistic Formulation}
\label{sec:autologistic}

The autologistic model was developed with a different perspective: What form must the MRF conditional distributions take to have exponential family form and still be faithful to the NUG, $\N$?
\citet{besag1974} provided a proof for a general exponential family and further gives the result for binary data that 
\begin{equation}
\label{eq:autologistic_full_cond}
q(y_i|\y_{\partial(i)}) = \frac{\exp\left(y_i\left(\alpha + \psi\sum_{j\in\partial(i)}y_j\right)\right)}{1 +\exp\left(\alpha + \psi\sum_{j\in\partial(i)}y_j\right)},
\end{equation}
for $y_i\in\{0,1\}$.
This formulation leads to the same analytic form for the pairwise clique potential as the physics-Ising formulation, with the key difference in the coding of the binary observations, resulting in a model with a distinct encoding of spatial dependence.
We give the definition of the autologistic formulation below.  
\begin{definition}
    For $\y\in\{0,1\}^n$, $g(y_i)=y_i$ and $g(y_i,y_j)=y_iy_j$, an MRF has an autologistic formulation if
    \begin{equation*}
    \label{eq:autologistic}
           q(\y|\xibf) = \frac{1}{Z(\xibf)} \exp\left(\alpha\sum_i y_i + \psi\sum_{i\sim j}y_i y_j \right).
    \end{equation*}
\end{definition}
In the physics-Ising formulation, it is natural to talk about up and down spins, while, here, a zero-one coding lends itself well to a black and white image analogy.
We will use this black and white picture terminology throughout the remainder of the paper, where a ``black'' unit corresponds to $y_i=1$ and ``white'' -- $y_i=0$. 
In the autologistic formulation, $T_1(\y)\in\{t:t\in\mathbb{N},t\le|V|\}$ and $T_2(\y)\in\{t:t\in\mathbb{N},t\le|E|\}$.
The statistic $T_1$ counts the number of black areal units and $T_2$ counts the number of black and black matches only. 
The difference in sufficient statistics in the symmetric physics-Ising formulation and asymmetric autologistic model has important consequences.   
In the autologistic formulations, for two configurations where $\y^*$ has one additional black areal unit than $\y$ but $T_2(\y^*)=T_2(\y)$, the log-odds of $\y^*$ to $\y$ is equal to $\alpha$. 
Likewise, assuming $T_1(\y^*)=T_1(\y)$, but incrementing the number of black and black matches in $\y^*$ by one gives the log-odds of $\y^*$ to $\y$ to be $\psi$. 
In contrast, there is no difference in the log-odds when the number of white-white matches increase by one and $T_1(\y^*)=T_1(\y)$.
The asymmetry of $\psi$ is also clear in terms of the log-odds of the full-conditional distribution of variable $y_i$.
Every neighbor of areal unit $i$ that has the value one increases the log-odds of $y_i$ being one given its neighbors by $\psi$.
\citet{caragea_kaiser2009} propose a centered autologistic model to address this asymmetry. In Section Section~\ref{sec:covariate_external_field}, we use response functions to highlight an undesirable property of this modification of the autologistic formulation.
We also note in Section~\ref{sec:phase_transition}, that unlike the other three presented presented, the autologistic model does not exhibit phase transition and discuss the implications thereof. 

\subsubsection{Ising Formulation}
\citet{potts1952} extended the physics-Ising formulation to categorical setting.  The 2-state version of the Potts formulation for binary data where $y_i$ is zero-one coded is commonly called the ``Ising model'' in the statistics literature. We follow this convention here and provide the following definition.   
\begin{definition}
 For $\y\in\{0,1\}^n$, $g(y_i)=y_i$ and $g(y_i,y_j)=I(y_i=y_j)$, an MRF has an Ising formulation
    \begin{equation*}
    \label{eq:ising}
           q(\y|\xibf) = \frac{1}{Z(\xibf)} \exp\left(\alpha\sum_i y_i + \psi\sum_{i\sim j}I(y_i=y_j) \right).
    \end{equation*}
\end{definition}
This is the formulation most commonly used when statisticians specify an Ising prior in a hierarchical model.
Indeed the physics-Ising formulation and Ising formulation are related by a factor of 2. 
For the Ising model we have $T_1(\y)\in\{t:t\in\mathbb{N},t\le|V|\}$ and $T_2(\y)\in\{t:t\in\mathbb{N},t\le|E|\}$, where again the sample space of $T_2$ is possibly a subset of the set given.
Holding $T_2$ constant, changing a single areal unit from white in $\y$ to black in $\y^*$ gives the log-odds equal to $\alpha$.
Thus $\alpha^{(I)} = 2\alpha^{(PI)}$, where superscript $(I)$ denotes the Ising formulation and $(PI)$ the physics-Ising.
Because $T_2$ counts the number of matches (both black-black and white-white), we also have $\psi^{(I)} = 2\psi^{(PI)}$.
As follows, holding $T_1$ constant and increasing the number of matches by one in $\y^*$ from $\y$ gives the log-odds equal to $\psi$. 
This can also be verified by the fact that $T_2^{(PI)}=2T_2^{(I)}-|\E|$. 
The full-conditionals for the Ising model are 
\begin{equation}
\label{eq:ising_full_cond}
q(y_i|\y_{\partial(i)}) = \frac{\exp\left(\alpha y_i + \psi\sum_{j\in\partial(i)}I(y_j=y_i)\right)}{\exp\left(\psi\sum_{j\in\partial(i)}I(y_j=0)\right) +\exp\left(\alpha + \psi\sum_{j\in\partial(i)}I(y_j=1)\right)},
\end{equation}
for $y_i\in\{0,1\}$.

\subsubsection{Potts Formulation}
\label{sec:standard_potts_formulation}
We now give the standard Potts formulation where $y_i$ can take on $k$ distinct categories $\{0,1,\dots,k-1\}$. A more flexible version of the Potts formulation for categorical data is given in Section~\ref{sec:flexible_potts_formulation}, with an extension for ordinal data stated in Section~\ref{sec:ordinal_potts_formulation}. Continuing with the image analogy, the $k$ categories are often called colors. 
The Potts formulation definition is as follows.
\begin{definition}
    For $\y\in\{0,1,\dots,k-1\}^n$, $g_l(y_i)=I(y_i=l)$ for $l\in\{0,1,\dots,k-1\}$ and $g(y_i,y_j)=I(y_i=y_j)$, an MRF has a Potts formulation
    \begin{equation*}
    \label{eq:potts-2}
           q(\y|\xibf) = \frac{1}{Z(\xibf)} \exp\left(\sum_{l=0}^{k-1}\alpha_l\sum_i I(y_i=l) + \psi\sum_{i\sim j}I(y_i=y_j) \right).
    \end{equation*}
\end{definition}
For identifiability, we set $\alpha_l=0$. Alternatively, we could impose the constraint $\sum_{l=0}^{k-1}\alpha_l=0$. 
The sufficient statistic for the vector $[\alpha_0,\dots,\alpha_{k-1}]'$ is the sum of each color $l$, and we define $T_{1l}(\y)=\sum_i I(y_i=l)$.
As before, holding $T_2$ constant, we have that $\alpha_l$ equals the log-odds of changing a single areal unit not of color $l$ to color $l$.
The interpretation of $\psi$ is the same as in the Ising formulation. 

\section{Response Functions}
\label{sec:response_functions}
In this section we introduce a new prior analysis tool, response functions, which are particularly well suited to characterize how different formulations of the MRF encode spatial dependence. 
Response functions, as a prior analysis technique, fit nicely into the Bayesian workflow of \citet{gelman_etal2020}.
Define $\mathbb{R}$ to be the set of real numbers.
Let $\Fcal:p \rightarrow \mathbb{R}$ be a functional mapping of the sampling distribution of $T(\y)$ induced by the sampling model, $p(\y|\thetabf)$, to a real-valued number, which we denote by $\Fcal_{\thetabf}(T(\y))$.
The key idea of a response function is to examine the effect of a single parameter, $\theta_i$, on the functional, $\Fcal$, of the sampling distribution of $T(\y)$, with respect to a sampling model, $p(\y|\thetabf)$, holding $\thetabf_{-i}$ constant.
Here we use $\thetabf_{-i}$ to denote the set/vector of parameters with the $i$th parameter removed.
Most commonly, we will examine the expected value and variance together as the functionals of the summary statistic sampling distribution with respect to $p(\y|\thetabf)$. 
The response functions for a collection of summary statistics provide useful insights into the behavior of the sampling model.
The definition of a response function is as follows. 
\begin{definition}
\label{defn:prf}
    For a statistic $T(\y)$, sampling model $p(\y|\thetabf)$, with parameters $\thetabf$, and functional $\Fcal$, the $\Fcal$-response function for $\theta_i$ is the functional of the sampling distribution of the summary statistic induced by $p(\y|\thetabf)$,
    $$
    h_{T(\y)}^\Fcal(\theta_i|\thetabf_{-i}) = \Fcal_{\thetabf}(T(\y)), 
    $$
    as a function of $\theta_i$.
\end{definition}

Later in Section~\ref{sec:phase_transition} we will examine the mean and variance response functions of $T_1(\y)$ and $T_2(\y)$ to understand the phenomena of phase transition exhibited in some formulations of the MRF. 
Rather than hold $\thetabf_{-i}$ constant, we can place a prior on $\thetabf$ and marginalize out $\thetabf_{-i}$ over the conditional prior $p(\thetabf_{-i}|\theta_i)$.
This yields a \textit{prior predictive response function}, defined below.

\begin{definition}
\label{defn:pprf}
    For a statistic $T(\y)$, sampling model $p(\y|\thetabf)$ with parameters $\thetabf$, conditional prior $p(\thetabf_{-i}|\theta_i)$, and functional $\Fcal$, the prior predictive response function of parameter $\theta_i$ is the functional of the distribution of the summary statistic,
    $$
    h_{T(\y)}^\Fcal(\theta_i) = \Fcal_{\theta_i}(T(\y)),
    $$
    with respect to the marginal sampling model $p(\y|\theta_i)=\int p(\y|\thetabf)p(\thetabf_{-i}|\theta_i) d\thetabf_{-i}$. 
\end{definition}

When the functional of the sampling distribution of $T(\y)$ is not known, we can estimate the response and prior predictive response functions through simulation.

\subsection{Monte Carlo Estimation}
Estimation of the response and prior predictive response functions is straight forward as long as we can obtain draws from $p(\y|\thetabf)$. For a review of sampling algorithms for MRFs see \citet{izenman2021}. 
To obtain an estimate of $h^{\Fcal}_{T(\y)}(\theta_i|\thetabf_{-i})$, fix $\thetabf_{-i}=\thetabf_{-i}^*$ and let $\w=[w_1,\dots,w_J]$ be sequence of points covering the range of $\theta_i$. For each $j=1\dots,J$, obtain $B$ samples from $p(\y|\theta_i=w_j,\thetabf_{-i}^*)$ and then estimate the response function at each $w_j$ as
$$
\hat h^{\Fcal}_{T(y)}(\theta_i=w_j|\thetabf_{-i}^*)= f(\{T(\y_{jb})\}_{b=1}^B),
$$
where $\y_{jb}$ denotes the $b$th sample when $\theta_i=w_j$ and $f$ is the function that can be applied to an empirical distribution (i.e., the set of draws $\{T(\y_{jb})\}_{b=1}^B$, that corresponds to the functional $\Fcal$).
For example, when $\Fcal$ is the expectation operator, we have
$$
f(\{T(\y_{jb})\}_{b=1}^B) = \frac{1}{B}\sum_{b=1}^{B} T(\y_{jb})
$$
and for the variance we have
$$
f(\{T(\y_{jb})\}_{b=1}^B)=\frac{1}{B-1}\sum_{b=1}^{B}\left(T(\y_{jb})-\frac{1}{B}\sum_{b=1}^{B} T(\y_{jb})\right)^2.
$$
The same $J\times B$ samples can be used to estimate other summary statistics $T(\y)$ and functionals of interest.

The prior predictive response function Monte Carlo estimate can be optained in similar manner. For each $j=1,\dots,J$, we first obtain $B$ draws of $\thetabf_{-i}$ from $p(\thetabf_{-i}|\theta_i)$ which reduces to $p(\thetabf_{-i})$ if $\thetabf_{-i}$ and $\theta_i$ are \textit{a priori} independent. Then for $b=1,\dots,B$, we obtain a draw from $p(\y|\theta_i=w_j,\thetabf_{-i,b})$ and estimate
$$
\hat h^{\Fcal}_{T(y)}(\theta_i=w_j)= \frac{1}{B}\sum_{b=1}^{B}f(T(\y_{jb})),
$$
where $\y_{jb}$ are draws from the distribution $p(\y|\theta_i=w_j)$. 
If $\w$ is a sufficiently fine grid and $B$ is sufficiently large such that the Monte Carlo error of the estimate is negligible, the response function can be approximated with line segments between $(w_j,\hat h^{\Fcal}_{T(y)}(\theta_i=w_j|\thetabf_{-i}^*))$ and $(w_{j+1}, \hat h^{\Fcal}_{T(y)}(\theta_i=w_{j+1}|\thetabf_{-i}^*))$ for $j=1,\dots,J-1$. 
Otherwise a smoothing technique can be used to estimate $\hat h$. 
A similar approach can be used to produce Monte Carlo estimates of the prior predictive response functions. 

\subsection{Connection to Thermodynamic Quantities}
The prior analysis tool of response functions was directly inspired by the well studied thermodynamic quantities of the physics-Ising model.
Previously, we mentioned that $-\sum_{c\in C} f_c(\y|\xibf)$ is interpreted as the potential function, or Hamiltonian, in statistical mechanics.
Configurations of the graph with low potential energy are assigned higher probabilities in the MRF.
In this setting, the physics-Ising model serves as a mathematical model for ferromagnetism.
In Section~\ref{sec:physics-ising}, we presented a parameterization of the physics-Ising model that would be more familiar to statisticians and analogous to the other formulations of MRFs presented in Section~\ref{sec:discrete_mrfs}.
For a nice introduction to the more traditional parameterization used by physicists, see \citet{cipra1987}.

In describing the behavior of atoms on a lattice, physicists sought to understand how $\psi$ (in our formulation, a linear reparameterization of the inverse temperature) affects the magnetic behavior of the atoms.
The behavior of the system is summarized through thermodynamic quantities, which can often be expressed as functions of the log partition function.
For example, $\EE_{\xibf}(\T_1(\y)) = \frac{\partial A(\xibf)}{\partial \alpha}$, is the magnetization of the system for the physics-Ising model. 
Importantly, the thermodynamic quantities have real world interpretations for the ferromagnetic behavior of the atoms in the lattice. Thus physicists study thermodynamic quantities as a function of the inverse temperature and/or external field. 
In some cases, the function has an analytic form, while in other cases it is studied through simulations.
This notion of studying functional summaries of probability distributions, originally applied to describing physical system, is the exact same approach as our response function which we generalize as a prior analysis tool to help understand the sampling model.
In the following section, we will show how the response functions highlight the behavior of phase transition in some formulations of the MRF. 

\section{Phase Transition}
\label{sec:phase_transition}
As alluded throughout the paper, a key property of the Ising and Potts models is the phenomenon of phase transition.
Physicists study the phenomenon through thermodynamic quantities; in statistical modeling, response functions help to understand the encoding of spatial dependence in an MRF and to elicit priors for the spatial dependence parameter.
We will explore implications of phase transition on simulating configurations from an MRF in Section~\ref{sec:critical_slowing}. 
Because of the importance of phase transition in the context of physics, the term is often included in the statistics literature, but the definition sometimes remains elusive.
A simple definition may seem tautological: phase transition is the change from one state to another state.
A substance going from a liquid state to a solid state is a phase transition.
In this freezing example, there is a critical temperature at which the substance undergoes a shift in the order of the system.
Above the critical temperature the molecules interact in a way such that the substance exhibits the characteristics of a liquid and below the critical temperature the substance becomes solid.
A sharp dichotomy in the order of the system exists at the critical temperature.
For the Potts and Ising formulations of the MRF (and physics-Ising), there is a distinct critical value, $\psi^*$, below and above which realizations from the MRF are markedly different. 
We now give two statistical interpretations of phase transition for discrete MRFs. 

\subsection{Statistical Interpretation}
In phase transition, the change in the system is marked by a discontinuity in the order of the system.
Physicists hypothesized and later proved that this discontinuity would appear in the thermodynamic quantities that describe the behavior of a ferromagnetic system. 
Following \citet{stoehr2017}, we give an explicit statistical interpretation of phase transition in terms of sufficient statistics, from which the thermodynamic quantities are also derived.
Before we explain the discontinuity, we provide a definition: We call two configurations, $\y^*$ and $\y$, with a zero-one coding \textit{complementary} if $\y^* = 1-\y$.
Now for illustration, assume that $\alpha=0$.
It is clear for the Ising formulation that $q(\y|\xibf)=q(1-\y|\xibf)$, as $g(y_i,y_j)=I(y_i=y_j)$ is symmetric.
Then the distribution of $T_1(\y)/n$ is symmetric and centered around $1/2$ as complementary configurations have the same probability and $T_1(\y)/n = 1-T_1(\y^*)/n$. 
It follows that the expected value of $T_1(\y)/n$ is equal to $1/2$.
Recall from Section~\ref{sec:exp_fam} that $\EE(T_1(\y))=\frac{\partial A(\xibf)}{\partial \alpha}$.
This result is true for all finite lattices and all values of $\psi$ in the Ising formulation with $\alpha=0$. 
Interestingly, as we allow $n\to\infty$ and $\alpha\to0$, the same result does not hold for all values of $\psi$.
For values of the spatial dependence parameter below the critical value, $\psi<\psi^*$, we have
$$
\lim\limits_{\alpha\to0}\lim\limits_{n\to\infty}\frac{1}{n}\frac{\partial A(\xibf)}{\partial \alpha} = \frac{1}{2}.
$$
As the external field is taken away ($\alpha \rightarrow 0$), the probability distribution will again assign equal probabilities to complementary configurations in the limit as $n$ goes to infinity.
For $\psi>\psi^*$, we have instead that 
$$
\lim\limits_{\alpha\to0^+}\lim\limits_{n\to\infty}\frac{1}{n}\frac{\partial A(\xibf)}{\partial \alpha} > \frac{1}{2}\;\; \text{and}\;\; \lim\limits_{\alpha\to0^-}\lim\limits_{n\to\infty}\frac{1}{n}\frac{\partial A(\xibf)}{\partial \alpha} < \frac{1}{2},
$$
where $\rightarrow0^+$ denotes the one-sided limit from above and $\rightarrow0^-$ the limit from below. 
The residual effect of the external field persists in the infinite system for values of $\psi$ above the critical value.
(See Figure 5 in \citet{cipra1987}'s review of the physics-Ising formulation.)
That is, the expected value of $T_1(\y)/n$ will still favor the original pull of the external field even as $\alpha\to0$. 
As noted above, for the physics-Ising formulation, $\frac{1}{n}\frac{\partial A(\xibf)}{\partial \alpha}$ can be conceptualized as the magnetization of the system.  
\citet{peierls1936} was the first to show that phase transition exhibited through ``spontaneous" magnetization does indeed exist for a two-dimensional lattice. (See \citet{duminil-copin2022} for a historical overview of the physics-Ising formulation and key developments.)

Later, \citet{kramers_wannier1941} hypothesized that the specific heat, which is a function of $\frac{\partial^2A(\xibf)}{\partial \psi^2}$, would show a discontinuity at $\psi^*$ when $\alpha=0$.
\citet{stoehr2017} succinctly expresses this discontinuity in terms of the variance of the pairwise sufficient statistic
$$
\underset{\psi\to\psi^*}{\lim}\lim\limits_{n\to\infty} \var(T_2(\y))=\infty.
$$
It is important to remember that phase transition only exists for the infinite lattice and that the discontinuities expressed in the above equations are smoothed over for finite lattices, that is, the limits exist (i.e., the limits are finite).
Even though phase transition technically only exists on a infinite lattice, the phenomenon and its effects are still evident on finite lattices. 
On a finite lattice, we see a sharp rise in the variance of $T_2(\y)$ as $\psi$ approaches the critical value.
For first-order lattices the critical value is known for the Ising and Potts model with $\psi^*=\log(1+\sqrt{k})$, where $k$ is the number of states. 

\begin{figure}[ht!]
    \centering
    \includegraphics[width=0.75\linewidth]{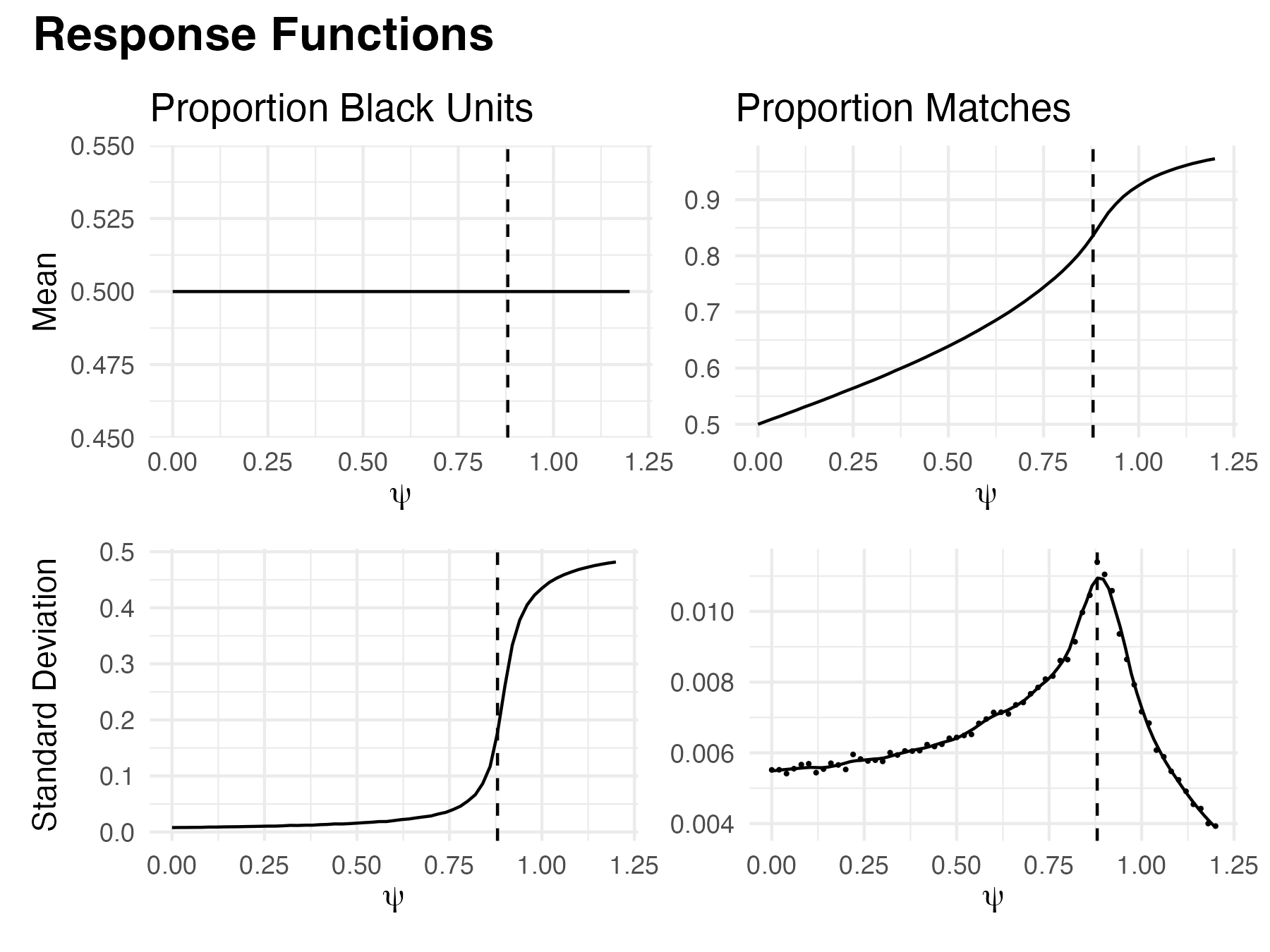}
    \caption{Monte Carlo estimates of the response functions for an MRF with an Ising formulation and no external field: the mean and variance of the distribution of the proportion of black units (left column) and the proportion of matches (right column). The dashed line marks the known critical value of $\psi\approx 0.88$.}
    \label{fig:rf_ising_alpha_0}
\end{figure}

We can use response functions to observe this phenomenon in a finite lattice.
Figure~\ref{fig:rf_ising_alpha_0} shows the mean and standard deviation response functions for the proportion of black units and for the proportion of matches in a $64\times64$ first-order lattice.
To estimate the response functions, we sampled 2,000 draws from the MRF for $\psi=0.0,0.02,\dots,1.2$.
A draw was obtained by performing 400 iterations of the Swendsen-Wang Algorithm \citep{swendsen_wang1987}, which converges faster than single vertex updates when $\psi$ is near the critical value (see Section~\ref{sec:critical_slowing}). 
For Monte Carlo estimates of response functions that did not render as smooth, we used the \texttt{geom\_smooth} function from \texttt{ggplot2} to smooth over the Monte Carlo error, which performs LOESS regression \citep{r_tidyverse2019}. Both the individual Monte Carlo estimates along with the smooth curve, if necessary, are shown in Figure~\ref{fig:rf_ising_alpha_0}.
The vertical dashed line marks the critical value of $\psi^*\approx.88$.
As expected the standard deviation of the proportion of matches rises sharply around the critical value.
This boundary marks the point at which the distribution of the proportion of black units shifts from unimodal to bimodal. 
After the critical value, we also see a sharp rise in the standard deviation of the proportion of black units, which corresponds to the bimodal distribution of nearly all black or nearly all white configurations. 

\subsection{Choice of Prior for the Spatial Dependence Parameters}
In Bayesian models with MRF components, a uniform prior on the interval $[0,\psi_{max})$ is often used for $\psi$ .
\citet{moores_etal2020} suggest selecting $\psi_{max}$ above $\psi^*$ to permit ``ordered'' configurations in the prior space of the MRF. 
In the context of spatial statistics, it is tempting to think of ``ordered'' as meaning configurations with spatial clustering, but realizations of an MRF will exhibit spatial clustering for $\psi<\psi^*$.
Because phase transition is not a sharp discontinuity in finite graphs, realizations of an Ising or Potts MRF for $\psi$ slightly above $\psi^*$ will not be configurations all of a single color, but will quickly be dominated completely by a single color as $\psi$ increases above $\psi^*$.
In other cases, a justification for a particular $\psi_{max}$ is not provided \citep{moller_etal2006, everitt2012bayesian, lyne2015russian}.
In the spirit of response functions, others use simulation studies to examine the proportion of matches in an Ising or Potts prior to determine where phase transition occurs in order to set $\psi_{max}=\psi^*$ \citep{green_richardson2002, cucala2009bayesian}.
In the authors' opinion, this is reasonable approach as practical applications do not exhibit random fields dominated by a single color.
Because of the information provided by observed data, \citet{moller_etal2006} notes that the maximum value for a uniform prior may not matter (as long as $\psi_{max}>\psi^*$) so a prior with nonnegative support (i.e. $\psi \in [0,\infty))$ could be used instead. 

\subsection{Critical Slowing Down}
\label{sec:critical_slowing}
The most consequential impact of phase transition on inference in a MRF is that of \textit{critical slowing down} for MRF samplers as $\psi\rightarrow\psi^*$.
We do provide a detailed review of sampling methods for MRFs; for such a review, we refer the reader to \citet{izenman2021}. 
Many inference techniques rely on sampling configurations from an MRF (see Section~\ref{sec:methods_fitting}) and response functions are also estimated Monte Carlo simulation.

Samplers that rely on single areal unit updates, such as the Gibbs sampler \citep{geman_geman1984} and coupling from the past \citep{propp_wilson1996} rely on performing a sufficient number of iterations, $M$, of individual vertex updates to the whole configuration so that the sampled configuration can be viewed as a sample from an MRF.
Here, one iteration of the algorithm is an update to every vertex in the configuration. 
In the case of coupling from the past, the algorithm guarantees an exact draw from an MRF as $M$ is adaptively increased until convergence is reached. 
The Gibbs sampler can be applied to all formulations (``Gibbs'' comes from Gibbs distributions, the title applied to MRFs in statistical mechanics which describe the energy of a system through pairwise interactions), whereas coupling from the past requires a partial ordering of configurations and can only be applied to the two state models and Potts model with $k\le 3$. 
For MRFs that exhibit phase transition, both of these algorithms with single areal unit updates suffer from a critical slowing down as $\psi \rightarrow {\psi^*}^-$ \citep{izenman2021}. 
Specifically, as $\psi \rightarrow {\psi^*}^-$, the number of iterations, $M$, increases for the two algorithms to converge, that is the rate of convergence is much slower for values of $\psi$ near the critical value and above. 
This is a lingering effect of phase transition on a finite lattice, as $\psi \rightarrow {\psi^*}^-$, the distribution of $T_1(\y)$ becomes increasingly flat until it becomes bimodal, in the Ising formulation, for $\psi>\psi^*$.
As a consequence, an algorithm that relies on single areal unit updates requires many more steps to explore the space efficiently.
In other words, it is quite difficult with single areal unit updates to go from a nearly all one color configuration to a nearly all the other color even though the two set of configurations are equally likely in the prior with no external field. 
As the autologistic formulation does not have a symmetric pairwise clique potential, it does not experience phase transition and samplers do not experience this critical slowing down. 

Two samplers were proposed to deal specifically with this critical slowing down: The Swendsen-Wang algorithm can be used to sample from a discrete MRF with any number of states \citep{swendsen_wang1987} and 
the Wolff algorithm efficiently samples configurations for binary MRFs \citep{wolff1989}. 
The Swendsen-Wang algorithm is known to behave poorly in the presence of an external field, so \citet{barbu2003graph} provide an updated Swendsen-Wang cuts algorithm that incorporates external field information.

\section{Parameterization of the External Field}
\label{sec:external_field}

In the simplest versions of the above formulations of MRFs, we specified the external field parameter, $\alpha$, to be constant across all areal units. 
A richer class of MRFs can be created when we provide a parameterization for $\alpha_i$ that depends on observed variables (i.e., covariates).
In the cases explored below, the external field is a parametric function of a vector of covariates, $\x_i$, associated with each areal unit, outcome variables $\y_i$, or a combination of covariates and outcome variables.

\subsection{External Field through Covariates}
\label{sec:covariate_external_field}
Consider a parameterization of the external field in an MRF as a linear combination of covariates with unknown coefficients, in a manner analogous to the multiple regression setting.
That is, we specify $\alpha_i = \x_i'\betabf$, where $\betabf$ is a vector of coefficients and $\x_i$ is a vector of the $i$th row of the matrix of covariates $\X$.
As a direct data model for binary outcomes we can write the formulation-agnostic probability model as
$$
q(\y|\betabf,\psi)=\frac{1}{Z(\betabf,\psi)}\exp\left(\sum_i y_i\left(\x_i'\betabf\right) + \psi\sum_{i\sim j} g(y_i,y_j) \right).
$$
In the Potts model with $k$ categories, we modify the external field to have category specific covariates, $\betabf_l$, for $l=0,\dots,k-1$, yielding the model
$$
q(\y|\{\betabf_l\}_{l=0}^{k-1},\psi)=\frac{1}{Z(\{\betabf_l\}_{l=0}^{k-1},\psi)}\exp\left(\sum_i \left(\sum_{l=0}^{k-1} I(y_i=l)\x_i'\betabf_l\right) + \psi\sum_{i\sim j} I(y_i=y_j) \right),
$$
where $\betabf_0=\0$ is set to a vector of zeros for identifiability.
When the spatial dependence parameter is zero, the full conditionals have the form of logistic regression likelihoods for the binary formulations (with a nonstandard parameterization when using the physics-Ising formulation) and multinomial regression likelihoods for the Potts formulation.
As we will explore below, when the spatial dependence parameter is not zero in the autologistic, Ising and Potts formulations, the marginal effect of a covariate on the probability that areal unit is equal to category $l$ holding all else constant still follows the same interpretation as in logistic and multinomial regression. 
That is, holding all else constant (neighboring variables and other covariates) for a one unit increase in covariate $x_j$, we expect the log odds that $y_i=l$ to increase by $\beta_j$ relative to $y_i\ne l$.
While the marginal interpretation is the same, the choice of clique function for spatial dependence does lead to distinct parameter effects on the probabilities of different configurations (and subsequent parameter interpretations) in the full model. 

While any of the MRF formulations can be used with a linear combination of covariates for the external field, historically, the autologistic formulation is the default direct data model for binary observations.
As discussed previously, the autologistic formulation only counts positive neighboring matches towards spatial dependence, which biases configurations to all ones. 
\citeauthor{caragea_kaiser2009}'s proposed solution -- the centered autologistic model --  to this potentially undesirable feature introduces a type of phase transition with undesirable consequences for parameter interpretation as we show in the next section. 

\subsubsection{Choices of Binary Direct Data Models with Covariates}
In this section, we compare the autologistic, centered autologistic, and Ising formulations as a direct data models, exploring model properties through prior predictive response functions and parameter interpretations through the full conditional distributions.
We begin with the historical default direct-data model for dependent binary outcomes associated with areal units: the autologistic model.
The autologistic model has full conditionals
$$
q(y_i|\y_{\partial(i)},\betabf,\psi) = \frac{\exp\left(y_i\left(\x_i'\beta + \psi\sum_{j\in\partial(i)}y_j\right)\right)}{1+\exp\left(\x_i'\beta + \psi\sum_{j\in\partial(i)}y_j\right)}.
$$
From the full conditionals, we see that any positive neighbor increases the conditional probability that $y_i=1$, while neighboring zeros have no effect on the conditional probability.
Even though the \textit{autocovariate}, a variable that is a function of the values of neighboring units (in this case, $\sum_{j\in\partial(i)}y_j$), of the autologistic model biases the conditional probability toward one over the independence model, the marginal effect of increasing a single covariate holding all else constant still has the same \textit{mathematical} interpretation as in logistic regression.
As a consequence of this bias, noted by \citet{caragea_kaiser2009}, the conditional expectation that $y_i=1$ in the autologistic model will always be greater than the conditional expectation under the independence model ($\psi=0$) whenever there is a neighbor with the value one (and $\psi > 0$).
The centered autologistic model addresses this discrepancy of conditional expectations by subtracting the expected value under the independence model from the autocovariate, a sort of centering for binary data, and writes the full conditional distributions as 
$$
q(y_i|\y_{\partial(i)},\betabf,\psi) \propto \exp\left(y_i\left(\x_i'\betabf + \psi\sum_{j\in\partial(i)}\left(y_j - \frac{\exp(\x_j'\betabf)}{1+\exp(\x_j'\beta)}\right)\right)\right).
$$
This centering was motivated by the idea that the autocovariate 
will increase the probability that $y_i=1$ when the number of neighbors that equal one 
is greater than the expected number of neighbors equal to one.
Interpreted another way, however, the autocovariate influences the probability of $y_i$ when the neighbors are misaligned with the expected value under the independence model.
Specifically, when $E(y_j)$ under independence is close to one, but $y_j$ is zero, this decreases the probability that $y_i=1$ and conversely, when $E(y_j)$ under independence is close to zero but $y_j$ is one, this increases the probability that $y_i=1$. 

Recall that an MRF can be specified through the full conditionals as long as we verify that it can be written equivalently in terms of clique potentials. 
Interestingly, to obtain the full conditional distribution above, the clique potential for the singleton requires the adjustment 
$$
f(y_i|\betabf,\psi)=y_i\left(\x_i'\betabf - \psi\sum_{j\in\partial(i)}\frac{\exp(\x_j'\betabf)}{1+\exp(\x_j'\betabf)}\right),
$$ while the pairwise clique function, $f(\y_i,\y_j|\psi)=\psi y_i y_j$, remains the same as in the autologistic formulation \citep{hughes_etal2011}. 
The goal of this model was to create ``interpretable'' model parameters, namely, interpretation of the external field through $\X\betabf$ and local spatial dependence through the parameter $\psi$.
This formulation does not appear to be more interpretable, in fact, the effect of the spatial dependence parameter is now more convoluted as it is required to show up in the singleton clique potential. 
While \citet{caragea_kaiser2009} perform a simulation study to show that conditional expectation under their centered autologistic is approximately equal to the conditional expectation under the independence model, both the single clique potential and full conditional distributions of the centered autologistic seem to indicate a misalignment with the external field, especially as $\psi$ increases.
We show that this misalignment is apparent through prior predictive response functions in the simulation below.



Instead of the centered autologistic model with nonintuitive clique potentials as a solution to the bias of the autologistic model, one need only choose a symmetric pairwise clique potential as in the Ising formulation. 
With the symmetric pairwise clique function, we obtain full conditionals
$$
q(y_i|\y_{\partial(i)},\betabf,\psi) = \frac{\exp\left(y_i(\x_i'\beta) + \psi\sum_{j\in\partial(i)}I(y_i=y_j)\right)}{\exp\left(\psi\sum_{j\in\partial(i)}I(y_j=0)\right)+\exp\left(\x_i'\beta + \psi\sum_{j\in\partial(i)}I(y_j=1)\right)},
$$
which increases the probability that $y_i=1$ when the majority of neighbors are one and likewise increases the probability of $y_i=0$ when the majority of neighbors are zero. 
This model does not have the same conditional expectation as the independence model, except when the neighboring values are equally distributed between one and zero, but this is in fact a desirable property and the point of including a spatial autocovariate. 
We now compare the autologistic, centered autologistic, and Ising formulations in the simulation study below. 

\subsubsection{Comparison of Binary Direct Data Models through Prior Predictive Response Functions}
To compare the effect of the spatial dependence parameter $\psi$ for the autologistic, centered autologistic, and Ising formulations of an MRF with an external field from covariates, we estimate a collection of response functions of the spatial dependence parameter $\psi$.
For our response functions, we examine the expected value and standard deviation of the proportion of black units, the expected value and standard deviation of the proportion of matches, the expected proportion of the dominant color, $T^{DC}(\y)=\max\left(\sum_i I(y_i=0),\sum_i I(y_i=1)\right)/n$, and the expected misclassification rate.
For the misclassification rate, we apply a simple decision rule to the expected value of each areal unit based on the external field in the independence model and compare the simulated MRF configuration to our external field classification. 
For each areal unit, the external field classification is
$$
a_i = I\left(\frac{\exp(\x'\betabf)}{1+\exp(\x'\betabf)} > 0.5 \right).
$$
The misclassification rate for a configuration $\y$ is
$$
T^{\mathrm{MR}}(\y) = 1 - \frac{1}{n} \sum_{i=1}^n I(y_i = a_i),
$$
one minus average agreement between the configuration and external field classification.

\begin{figure}[ht!]
    \centering
    \includegraphics[width=0.5\linewidth]{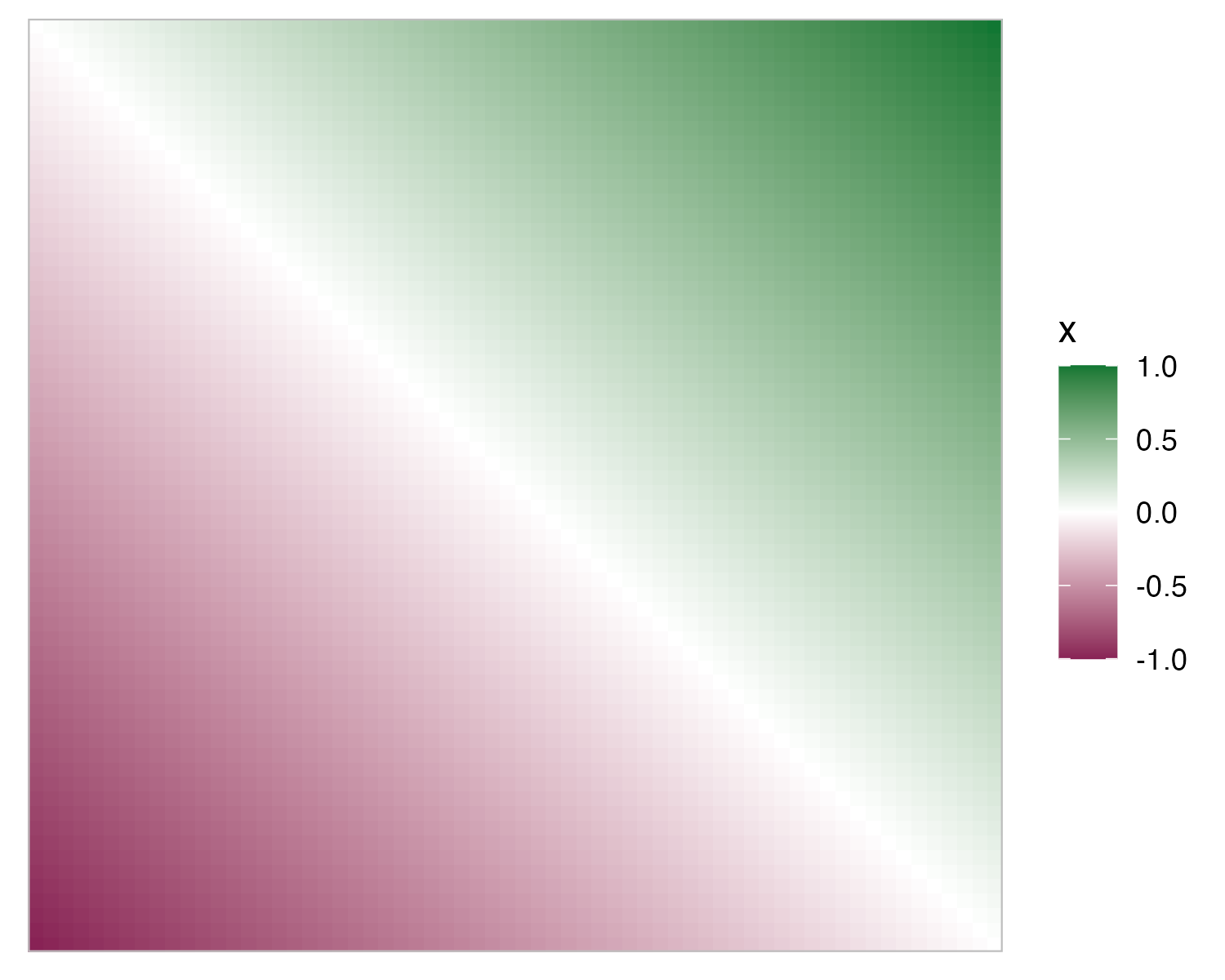}
    \caption{Plot of the covariate used in the response function and prior predictive response function simulation studies for binary MRF formulations with an external field parameterized as a linear function of the covariate.}
    \label{fig:covariates_sim_study}
\end{figure}

In our simulation study, following \citet{caragea_kaiser2009} set-up,  we construct a single covariate on a $64\times 64$ lattice by letting
$x_i=(r(i)+c(i)-n_r-1)/(n_r-1)$, where $r(i)$ returns the row of areal unit $i$, $c(i)$ returns the column, and $n_r=64$, the number of rows/columns. 
See Figure~\ref{fig:covariates_sim_study} for a visualization. 
We specify the external field as $\alpha_i = \beta_0 + x_i\beta_1$ and examine the prior predictive response functions of $\psi$ with independent standard normal priors on $\beta_0$ and $\beta_1$. 

To generate the prior predictive response functions, we first sample $\beta_0$ and $\beta_1$ from $\mathrm{N}(0,1)$ and then obtain a configuration of the MRF given $\beta_0$ and $\beta_1$.
For the Ising model, we perform 400 iterations of the Swendsen-Wang algorithm \citep{swendsen_wang1987} to obtain a configuration and for the autologistic and centered autologistic models obtain exact samples using the coupling from the past algorithm \citep{propp_wilson1996}. 
We follow this procedure 2,000 times for each value of  $\psi=0.0,0.03,\dots,1.7$ to obtain an estimate of the distribution of the proportion of black units, the proportion of matches, the dominant color, and the misclassification rate.

\begin{figure}[ht!]
    \centering
    \includegraphics[width=0.8\linewidth]{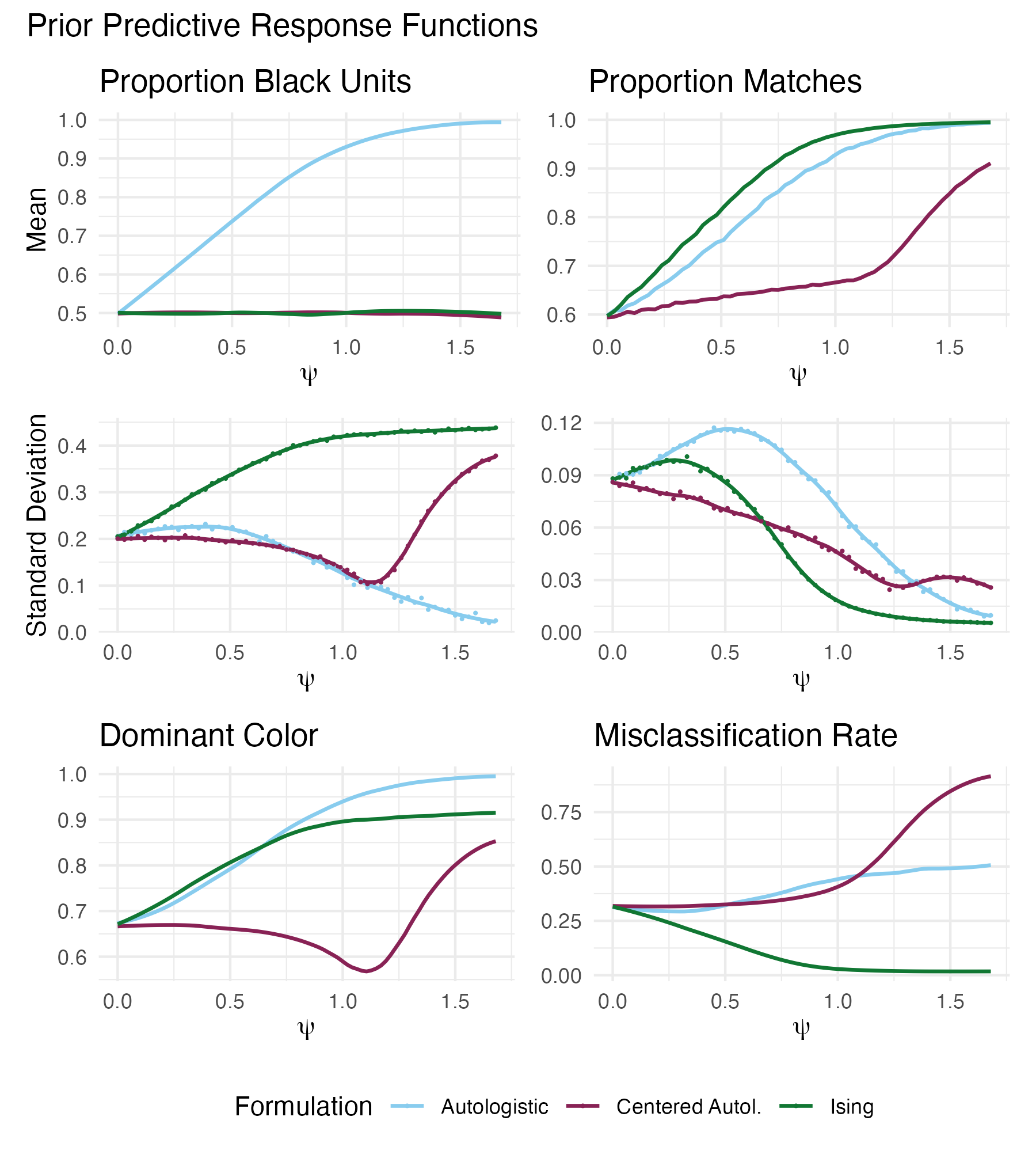}
    \caption{Plotted are the estimated prior predictive response functions for the autologistic, centered autologistic, and Ising models with $\beta_1$ and $\beta_0$ marginalized out over the prior $\mathrm{N}(0,1)$. The covariates are a planar gradient such that $x_i=(r(i)+c(i)-n_r-1)/(n_r-1)$. The top four plots show the estimated response functions corresponding to the mean and standard deviation of the proportion of black units and the proportion of matching neighbors. The bottom row shows the expected dominant color and expected misclassification rate.}
    \label{fig:pprf_int_norm}
\end{figure}

In Figure~\ref{fig:pprf_int_norm}, we show the prior predictive response functions for MRFs simulated with an external field and standard normal priors on the coefficients. 
In the top left panel we see that the estimated expected proportion of black units is $0.5$ across values of $\psi$ for the centered autologistic and Ising models, whereas the autologistic model favors configurations of all ones as $\psi$ increases.
The misalignment with the external field classification of the centered autologistic model is demonstrated in the bottom right panel.
Over the marginalized prior, the misclassification rate of the centered autologistic model never improves upon the misclassification rate under independence ($\psi=0$) and, in fact, begins to rise sharply after some critical value $\psi^*$, demonstrating greater misalignment with the external field for $\psi>\psi^*$, a type of phase transition.
The dominant color summary statistic also helps understand how the centered autologistic model encodes spatial dependence; it maintains the same expected dominant color as the independence model for $\psi$ approximately less than $.75$.
This is the stated goal of the centered autologistic model--to maintain approximate equivalence of the conditional expectation as compared to the independence model.
This is achieved, however, by discounting the influence of the external field and, for $\psi$ large enough, causes misalignment with the external field. 
The Ising model formulation, on the other hand, does not maintain the expected dominant color as the independence model.
The external field and pairwise clique potentials operate together to pull configurations towards the direction indicated by the external field.
As $\psi$ increases we obtain configurations that show greater alignment with the external field classification as the spatial dependence parameter influences areal units with weak external field signal to algin with values of neighboring units. 
Also of note, the centered autologistic model does not have a very high expected proportion of matches until phase transition (i.e., when misalignment with the external field occurs). 
This implies that the centered autologistic formulation is more restrictive \textit{a priori} on the degree of clustering in configurations than the Ising model. 
In conclusion, for binary models we recommend the Ising formulation, unless the bias of the autologistic model is a desired property and meaningful to the application. 

\subsection{An External Field through Hierarchical Modeling}
\label{sec:hierarchical_external_field}
Perhaps the most common application of Markov random fields is the hidden Markov random field (HMRF), where observations are modeled as conditionally independent given a spatially dependent categorical variable.
We do not give a complete review of the different modeling choices for HMRFs, but rather draw connections to the general MRF framework for discrete-valued variables, first by showing how the choice of observed variable distribution acts as an implicit external field. Then in Section~\ref{sec:pairwise_clique_potential}, we provide some extensions to the standard Potts model with different formulations of the pairwise clique function. 

Let the distribution of our observed variables given the latent discrete variable, sometimes called the \textit{emission distribution}, follow an exponential family
$$
p(y_i|z_i=l,\thetabf) = h(y_i)\exp(\etabf(\thetabf_{l})'\T(y_i) - A(\thetabf_{l})),
$$
where the complete set of parameters, $\thetabf=\{\thetabf_{l}\}_{l=0}^{k-1}$, are indexed by $z_i$ and where the vector-valued $\etabf$ and $\T$ and scalar-valued $h$ and $A$ are all known functions. 
Conditional on the observed $\y$, we again have an MRF distribution for $\z$,
$$
p(\z|\y,\thetabf, \psi) \propto \exp\left(\sum_{i=1}^n\sum_{l=0}^{k-1}I(z_i=l)\left( \etabf(\thetabf_{l})'\T(y_i) - A(\thetabf_{l})\right) + \sum_{i\sim j}f(z_i,z_j|\psi) \right),
$$
which is an exponential family MRF distributions with external field given by the specified outcome distribution.
For example, with a Potts formulation and Gaussian outcome distribution, we have singleton clique potential
$$
f(z_i|y_i,\mubf,\sigmabf)=I(z_i=l)\left(\frac{-(y_i-\mu_{l})^2}{2\sigma_{l}^2}-\log\sigma_{l}\right),
$$
for $l=0,\dots,k-1$, and pairwise clique potential $f(z_i,z_j|\psi)=\psi I(z_i=z_j)$.



\section{Extensions of the Pairwise Clique Potentials}
\label{sec:pairwise_clique_potential}
In this section, we provide two possible extensions to the standard Potts formulation described in Section~\ref{sec:standard_potts_formulation} and also introduce the modeling framework of conditional random fields.

\subsection{A More Flexible Potts Formulation}
\label{sec:flexible_potts_formulation}
The Potts model is the default specification for the prior on the categorical latent variable in an HMRF, although the single spatial dependence parameter may not be sufficiently expressive in some settings.  
The HMRF, the natural higher-dimensional extension of a hidden Markov model, typically has a single spatial dependence parameter, $\psi$, for the pairwise cliques that do not depend on the values of neighboring variables. 
In contrast,  a hidden Markov model has distinct transition probabilities from category $l$ to category $l'$ for $l,l'\in\{0,1,\dots,k-1\}$.
The analogous extension to the standard Potts formulation is to specify a pairwise clique function which takes into account alignment across all categorical pairs, not only matching neighbors.
Such a specification can be achieved with the pairwise clique potential
\begin{equation}
\label{eq:flexible_potts}
f(z_i,z_j|\psibf)=\sum_{r=0}^{k-1}\sum_{s=0}^{k-1}\psi_{rs}I(z_i=r)I(z_j=s),
\end{equation}
and the constraint that $\psi_{rs}=\psi_{sr}$.
The sufficient statistic for each $\psi_{rs}$ is simply the sum of neighboring pairs with one variable equal to $r$ and the other equal to $s$. 

\subsection{An Ordinal Potts Formulation}
\label{sec:ordinal_potts_formulation}
If the order of the categories is meaningful, the pairwise clique function can be modified to prefer neighboring values which are closer in value along the ordinal ranking.
\citet{feng_etal2012} provide a general formulation of the pairwise clique potential for an ordinal Potts model with
$$
f(z_i,z_j|\psibf)= \begin{cases}
    \psi_1 & \text{if }z_i=z_j,\\
    \psi_2 & \text{if }|z_i-z_j|=1,\\
    \psi_3 & \text{otherwise},
\end{cases}
$$
and $\psi_1\ge\psi_2\ge\psi_3$. In an application to MRI tissue classification, they let $\psi_1>0$, $\psi_2=0$, and $\psi_3<0$, which encourages neighboring matching pairs and penalizes neighbors that are farther than one ordinal category apart. This set-up has been dubbed the \textit{repulsion} Potts model. 

\subsection{Conditional Random Fields}
In the sections above, we showed that in a HMRF model, an MRF is a conjugate prior distribution for the latent variable when the outcome distribution is an exponential family. 
Stated differently, the full conditional posterior distribution, $q(\z|\y, \thetabf, \psi)$, is a Markov random field when $q(\z|\psi)$ is an MRF and $p(\y|\thetabf)$ follows an exponential family distribution.
In an HMRF, we assumed a parametric distribution for $p(\y|\thetabf)$. 
Conditional random fields, on the other hand, explicitly skip specifying the outcome distribution and define a MRF for $q(\z|\y, \xibf)$ directly, which allows for greater flexibility in formulating the dependence between the observed variables, $\y$, and latent variables, $\z$.
Here, the data generating process of $\y$ is unspecified.  Instead, the only assumption is that the latent and observed variables are mutually dependent. 
CRFs are often used to predict $\z$ given inputs $\y$. This requires that the model is trained (model parameters $\xibf$ are learned) for known $\z$ and $\y$. 
Then, after fitting a CRF and given new data, $\y^*$, the model can be used to predict the labels, $\z^*$ for the new inputs. 
\citet{lafferty2001conditional} originally proposed CRFs for labeled sequences, such as part of speech tagging and text processing, but CRFs have since been applied to many different domains \citep{yu2020comprehensive}.
Because we condition on $\y$ explicitly, we do not have to constrain the singleton clique functions to follow some parametric family form as in the HMRF, offering greater flexibility in the way we specify our Markov random field.
With explicit conditioning on $\y$ in our clique potentials, we have 
$$
q(\z|\y,\alphabf,\psibf) = \frac{1}{Z(\alphabf,\psibf)} \exp\left(\sum_i f(z_i|y_i,\alphabf) + \sum_{i\sim j} f(z_i,z_j|y_i,y_j,\psibf) \right),
$$
as yet another formulation of an MRF. 
As in the standard formulations of an MRF, the typical CRF decomposes the clique potentials as
$$
f(z_i|y_i,\alphabf)=\sum_u\alpha_ug_u(z_i,y_i),
$$ and 
$$
f(z_i,z_j|y_i,y_j,\psibf)=\sum_v\psi_vg_v(z_i,z_j,y_i,y_j),
$$ where each $g_u$ and $g_v$ returns a Boolean value for the different inputs of $\z$ and $\y$ to represent distinct features of the variables/vertices, and edges.
By constructing these Boolean feature functions for $z_i\in\{0,\dots,k-1\}$ and $y_i\in\{0,\dots,l-1\}$ as $g_{rs}(z_i,y_i)=I(z_i=r)I(y_i=s)$, for the singleton clique potential, we have
$$
f(z_i|y_i,\alphabf) = \sum_{r=0}^{k-1}\sum_{s=0}^{l-1} \alpha_{rs}I(z_i=r)I(y_i=s).
$$
The above formulation can equivalently be written as a covariate matrix with indicator variables for $y_i$ equal to category $s$ and distinct vector of coefficients for category $r$ of the latent variable.
Notice, that writing the Boolean features as indicator variables in a covariate matrix is equivalent to specifying the external field through a linear combination of covariates.
The key difference is not the formulation of the models, but rather the modeling scenario: CRFs can be viewed as a discriminative model, used to predict a latent variable given new discrete inputs, whereas direct data models seek to learn the relationship between covariates and the observed outcome.
While direct data models use covariates for the external field, CRFs further extend modeling flexibility by also allowing Boolean features to be constructed for the edges of a graph by considering pairs of latent and observed variables together through $g_v(z_i,z_j,y_i,y_j)$. 
This specification introduces dependence between $z_i$ and $\y_{\partial(i)}$ and typically, the pairwise clique is specified as the flexible Potts formulation above as in Equation~\ref{eq:flexible_potts}.
With such a specification, each $z_i$ is mutually dependent only with $y_i$. 

\section{Model Fitting}
\label{sec:methods_fitting}
The intractable normalizing constant is the main computational challenge of fitting an MRF model.
This is an issue whether inference on the spatial dependence parameter is desired or simply a point estimate of the latent spatial field is the main goal of analysis.
The property of phase transition remains important as the performance of different algorithms suffers near the critical value of the spatial dependence parameter. 
In this section, we provide a brief overview of the different techniques used to fit an MRF-based model.

The pseudo-likelihood, defined as the product of the full conditional distributions, was first proposed as an approximation to the MRF likelihood to facilitate maximum likelihood estimation \citep{besag1975}.
The pseudo-likelihood is known to perform poorly near the critical value, and the composite likelihood, a generalization, defined as the product of marginal and conditional likelihood components, can improve upon this inferences \citep{varin2011composite}. 
The pseudo-likelihood is sometimes used to replace the likelihood in Bayesian inference, but as the pseudo-likelihood is not a valid probability distribution \citep{friel_etal2009, cucala2009bayesian}, the probability guarantees of Bayes theorem do not hold \citep{carter2025mixture}.
\citet{matsubara_etal2024} propose a generalized Bayesian framework for intractable distributions which explicitly accounts the substitution of the likelihood with a generic loss function, and use the MRF as a motivating example. 

Approximate Bayesian computation, another likelihood-free alternative which compares summary statistics of simulated and observed data, has been applied to MRFs to avoid the cumbersome normalizing constant \citep{grelaud2009abc}.
The main computational hurdle is obtaining draws from the MRF distribution to compute the simulated summary statistics.
To improve upon this computational hurdle, \citet{moores_etal2020} introduce a surrogate modeling approach which incorporates precomputed response functions to facilitate scalable inference. 
Other Bayesian techniques avoid calculating the normalizing constant, but allow for exact inference with clever Metropolis-Hastings ratios that incorporate auxiliary data.
The auxiliary variable method \citep{moller_etal2006} and exchange algorithm \citep{murray_etal2006} require an exact sample from an MRF distribution, which can be acquired using coupling from the past \citep{propp_wilson1996}.
While the exact samples facilitate an exact MCMC algorithm, coupling from the past does not scale to large lattices (greater than 256 by 256) and additionally suffers from critical slowing down.

Other approaches seek to estimate the normalizing constant itself which can be done through path sampling or thermodynamic integration \citep{gelman1998simulating}. To decrease the computational cost of Monte Carlo estimation of the normalizing constant, \citep{green_richardson2002} discretize the prior for the spatial dependence parameter and precompute the estimates before performing full MCMC.
For small lattices, the normalizing constant can be computed exactly using a recursive algorithm \citep{reeves_pettitt2004}.  
Building upon this technique, this exact solution is used for smaller subsets of vertices of a larger lattice and results are combined to obtain an approximation of the normalizing constant for the whole lattice \citep{friel_etal2009, tjelmeland_austad2012}.
Partially-ordered Markov models \citep{cressie_davidson1998} have also served as the basis for constructing approximations to the normalizing constant \citep{austad_tjelmeland2017, chakraborty_etal2022}. 

Aside from inference on the parameters that govern the MRF, often a main inferential goal is estimation of the spatial field when it is not directly observed. 
Expectation-maximization (EM) type algorithms are often used, though the choice of likelihood approximation is still needed for updating of the spatial dependence parameter in the maximization step.
Variational techniques for approximations to the likelihood are often used to facilitate maximum a posteriori (MAP) estimates of the latent field \citep{jordan1999introduction, chatzis_tsechpenakis2010}.
The iterative conditional modes algorithm performs greedy coordinate updates using the full conditional distributions and is prone to converging to poor local minima \citep{besag1986}.
Simulated annealing introduces temperature-based exploration to escape local traps at greater computational cost \citep{geman_geman1984}.
Graph-cut algorithms such as the $\alpha$-expansion and $\alpha\beta$-swap procedures yield globally optimal (binary) or near-optimal (categorical) MAP solutions \citep{boykov2001interactive, zabih2004spatially}.


\section{Discussion}
\label{sec:discussion}

This review has provides a focused overview of Markov random fields (MRFs), primarily for binary data with connections drawn to more complex categorical modeling scenarios. 
We examined core structural properties such as factorization, neighborhood dependence, and phase transitions, and introduced response functions as a prior analysis tool to interpret how model parameters influence marginal and joint behavior. 
To illustrate these ideas in practice, we presented a case study of binary MRF formulations with covariates, demonstrating how the different direct-data model parameterizations encode dependence.

Our emphasis on binary MRFs was motivated by their foundational role in spatial statistics and probabilistic modeling.
Despite the analytic simple formulations of binary MRFs, key principles often remain ambiguous, which we sought to make clear in this review. 
While the binary and simple Potts models captures a wide range of modeling scenarios, many real-world applications rely on more complex or higher-dimensional extensions.
Notably, the broader literature on MRFs is vast and includes extensive developments in image analysis, where discrete-valued MRFs are used for texture modeling, denoising, and segmentation tasks \citep{li2009markov}.
Conditional random fields (CRFs), in particular, have emerged as a flexible alternative for discriminative modeling in computer vision and natural language processing \citep{sutton2012introduction}.
The principles and prior analysis tools highlighted in this paper provide a foundation for understanding more complex MRF distributions. 

While not a focus of this paper, graphical model selection and network learning share many principles with discrete MRFs, where the goal is to infer the underlying dependence structure from data.
In such contexts, the graph itself becomes the primary inferential target, and learning sparse, interpretable dependency structures often relies on penalized likelihoods \citep{meinshausen2006high, ravikumar2010high, drton2017structure}.
The interplay between structure learning and MRF-based modeling highlights the broader utility of graphical models in uncovering hidden structure in multivariate systems.

\bigskip
\begin{center}
{\large\bf SUPPLEMENTARY MATERIAL}
\end{center}

\begin{description}

\item[Supplement 1:] Code to reproduce the simulation studies is available in the zip folder.  

\end{description}

\bibliographystyle{abbrvnat}
\bibliography{references}

\end{document}